\documentclass[aps,prd,draft,showpacs]{revtex4}

\begin{document}

\title{Renormalization Group Running in the Symmetric and the Broken Symmetry Phases of the $R_{\xi } $
and the $\overline{R_{\xi }}$ Gauges}
\author{Chungku Kim}
\date{\today}

\begin{abstract}
We investigate the renormalization group (RG) running of the effective potential and the pole mass
in the broken symmetry phase of the $R_{\xi }$ and the $\overline{R_{\xi }}$ gauges
which have different RG running for the effective potential in the symmetric
phase and show that if the vacuum expectation value (VEV) is expressed as a function of the other
parameters of the theory by solving the minimization condition, then the
effective potential in the broken symmetry phase in both gauges satisfies the
same RG equation as one in the symmetric phase of the $\overline{R_{\xi }}$
gauge. The pole masses in the broken symmetry phase of both gauges are RG
invariant with respect to the RG funtions of the symmetric phase and are
shown to be the same at one-loop order.
\end{abstract}


\maketitle

\email{kimck@kmu.ac.kr}
\affiliation{Department of Physics, Keimyung University, Daegu 704-701, Korea}



\smallskip The $R_{\xi }$ gauge is widely used in gauge theory with spontaneous
symmetry breaking due to the fact that the mixing between the gauge field
and the Goldstone boson field in the kinetic term of the Lagrangian is
absent in the broken symmetry phase \cite{R-ksi}. Because the running mass of the
particles in the broken symmetry phase is given by the multiplication of the
coupling constants and the vacuum expectation value (VEV), the
renormalization group (RG) behavior of the VEV is very important. Recently,
the renormalization of the VEV in the $R_{\xi }$ gauge was investigated and the
resulting gamma function of the VEV turned out to be different from that of
the scalar field $(\gamma _{v}\neq \gamma _{\phi })$ [2,3]. 
This is the consequence of the fact that the $R_{\xi }$ gauge has a tadpole
divergence in the symmetric phase, and as a result, the scalar field needs
both multiplicative and additive renormalization [4,5]. 
This fact, as well as the fact that the Lagrangian depends on the VEV in
the symmetric phase causes a violation of the Higgs-boson low-energy
theorem \cite{Pilaft}. In order to avoid this problem, the non-linear $%
R_{\xi /\sigma }$ gauge was investigated [6,7]. 
Recently, it was shown that if the symmetric phase of the Lagrangian did not contain
the VEV, $\gamma _{v}$ coincides with $\gamma _{\phi }$ \ and the identity [8]
\begin{equation}
\xi \frac{\partial v}{\partial \xi }=C_{\xi }(v),
\end{equation}
held, where $C_{\xi }(v)$ was the function given by Nielsen [9]
and that if the symmetric phase of the Lagrangian depended on the VEV, which is the
case of the $R_{\xi }$ gauge, this identity should be modified \cite{kim2}.
Moreover, because most RG functions are calculated in the symmetric
phase of the MS\ Lagrangian, many attempts have been made to relate the RG
functions in the symmetric phase and those in the broken symmetry phase \cite
{Jegerlehner} in different renormalization schemes.

In this paper, we will investigate the RG behavior of the effective
potential and the pole mass in the broken symmetry phase of the $R_{\xi }$
and the $\overline{R_{\xi }}$ gauges, which is the $\sigma =1$ case of the $R_{\xi
/\sigma }$ gauge. For simplicity, we will consider the case of the Abelian
HIggs model with the Lagrangian density 
\begin{eqnarray}
L_{SYM}(\Phi _{1},\Phi _{2},A_{\mu }) &=&\frac{1}{4}F_{\mu \nu }F_{\mu \nu }+%
\frac{1}{2}(\partial _{\mu }\Phi _{1}+gA_{\mu }\Phi _{2})^{2}+\frac{1}{2}%
(\partial _{\mu }\Phi _{2}-gA_{\mu }\Phi _{1})^{2}+\frac{1}{2}m^{2}(\Phi
_{1}^{2}+\Phi _{2}^{2})  \nonumber \\
&&+\frac{\lambda }{24}(\Phi _{1}^{2}+\Phi _{2}^{2})^{2}+\frac{1}{2\xi }%
f(\Phi _{1},\Phi _{2},A_{\mu })^{2}+\overline{c}\frac{\delta f(A_{\mu
}^{\theta },\Phi _{2}^{\theta })}{\delta \theta }c+counter\text{ }terms,
\end{eqnarray}
where 
\begin{equation}
F_{\mu \nu }=\partial _{\mu }A_{\nu }-\partial _{\upsilon }A_{\mu },
\end{equation}
and $f(\Phi _{1},\Phi _{2},A_{\mu })$ is the gauge fixing function.

In the case of RG running for $\overline{R_{\xi }}$ gauge fixing in the broken
symmetry phase, the gauge fixing function is given by

\begin{equation}
f(\Phi _{1},\Phi _{2},A_{\mu })=\partial _{\mu }A_{\mu }-\xi g\Phi _{1}\Phi
_{2}.
\end{equation}
Because no tadpole divergence is possible in the symmetric phase, the scalar
fields are renormalized multiplicatively as usual, and the corresponding RG
equation for the renormalized effective action in the symmetric phase $%
\Gamma _{SYM}(\{c_{i}\},\phi )$ is given as

\begin{equation}
D\Gamma _{SYM}(\{c_{i}\},\phi )+\gamma _{\phi }\phi \frac{\partial \Gamma
_{SYM}(\{c_{i}\},\phi )}{\partial \phi }=0,
\end{equation}
where $\{c_{i}\}$ is the parameter set containing $\mu ,g,\lambda ,m^{2}$
and $\xi $, $\phi $ is the classical field of the scalar field $\Phi _{1}$,
and the operator $D$ is defined by 
\begin{equation}
D=\mu \frac{\partial }{\partial \mu }+\beta _{g}\frac{\partial }{\partial g}%
+\beta _{\lambda }\frac{\partial }{\partial \lambda }+\beta _{m^{2}}\frac{%
\partial }{\partial m^{2}}+\beta _{\xi }\frac{\partial }{\partial \xi }.
\end{equation}
Because $\Gamma _{SYM}(\Phi _{1},\Phi _{2},A_{\mu })$ does not depend on the VEV
in this gauge, the effective action in the broken symmetry phase $\Gamma
_{BS}(\{c_{i}\},\phi ,v)$ can be obtained as 
\begin{equation}
\Gamma _{BS}(\{c_{i}\},\phi ,v)=\Gamma _{SYM}(\{c_{i}\},\phi +v),
\end{equation}
where the VEV can be obtained from the minimization condition 
\begin{equation}
\left[ \frac{\partial V_{SYM}(\phi )}{\partial \phi }\right] _{\phi =v}=0,
\end{equation}
with the renormalized effective potential in the symmetric phase $V_{SYM}$
being obtained from $\Gamma _{SYM}$ by taking the classical field $\phi $ as a
constant field. By applying $D$ defined in Eq. (6) to this equation, we can
obtain $Dv=\gamma _{\phi }v$, which means that $\gamma _{v}=\gamma _{\phi }$  \cite{kim2}.
Then, by applying $D$ to the effective action in the broken
symmetry phase $\Gamma _{BS}(\{c_{i}\},\phi ,v)$ and by using Eq. (5), we
obtain 
\begin{eqnarray}
&&D\Gamma _{BS}(\{c_{i}\},\phi ,v)=D\Gamma _{SYM}(\{c_{i}\},\phi +v) 
\nonumber \\
&=&-\gamma _{\phi }(\phi +v)\frac{\partial \Gamma _{SYM}(\{c_{i}\},\phi +v)}{%
\partial \phi }+(Dv)\frac{\partial \Gamma _{SYM}(\{c_{i}\},\phi +v)}{%
\partial v}=-\gamma _{\phi }\phi \frac{\partial \Gamma _{BS}(\{c_{i}\},\phi
,v)}{\partial \phi }.
\end{eqnarray}
This means that the effective action in the broken symmetry phase $\Gamma
_{BS}(\{c_{i}\},\phi ,v)$ satisfies the same RG equation as that of the
effective action in the symmetric phase $\Gamma _{SYM}(\{c_{i}\},\phi )$ if
we make the substitution
\begin{equation}
v=v(\{c_{i}\}),
\end{equation}
as determined from the minimization condition given in Eq. (8) for the VEV in $%
\Gamma _{BS}(\{c_{i}\},\phi ,v)$.

In the case of RG running for $R_{\xi }$ gauge fixing in the broken symmetry
phase, the gauge fixing function is given by

\begin{equation}
f(\Phi _{1},\Phi _{2},A_{\mu })=\partial _{\mu }A_{\mu }-u\xi g\Phi _{2}.
\end{equation}
In this gauge, the tadpole divergence occurs in the symmetric phase[4,5]
Hence we need not only the multiplicative but also
the additive renormalization for the scalar fields as 
\begin{equation}
\phi _{B}=\sqrt{Z_{\phi }}(\phi +\frac{1}{2}u\text{ }\delta \widehat{Z}),
\end{equation}
which gives 
\begin{equation}
\mu \frac{\partial \phi }{\partial \mu }=\gamma _{\phi }\phi +\widehat{%
\gamma }\text{ }u,
\end{equation}
where $\widehat{\gamma }$ is the $O(\varepsilon ^{0})$ term of $\frac{1}{2}$ 
$\mu \frac{\partial \delta \widehat{Z}}{\partial \mu }.$ The resulting RG
equation in the symmetric phase $\Gamma _{SYM}(\{c_{i}\},u,\phi )$ becomes \cite{kim2} 
\begin{equation}
DV_{SYM}(\{c_{i}\},u,\phi )+(\gamma _{\phi }\phi +\widehat{\gamma }\text{ }%
u-C_{u}\gamma _{u})\frac{\partial V_{SYM}(\{c_{i}\},u,\phi )}{\partial \phi }%
=0,
\end{equation}
where $C_{u}$ is the function appearing in the Nielsen identity for the
gauge parameter $u$ as 
\begin{equation}
u\frac{\partial V_{SYM}(\{c_{i}\},u,\phi )}{\partial u}+C_{u}(\phi )\frac{%
\partial V_{SYM}(\{c_{i}\},u,\phi )}{\partial \phi }=0
\end{equation}
and 
\begin{equation}
\gamma _{u}=\frac{\mu }{u}\frac{\partial u}{\partial \mu }.
\end{equation}
Because the parameter $u$ of the gauge fixing function given in Eq. (11) should
be identified as the VEV $v$ in order to remove the mixing term between the gauge
field $A_{\mu }$ and the Goldstone field $\Phi _{2}$ in the kinetic part of
the Lagrangian, the effective action in the broken symmetry phase $\Gamma
_{BS}(\{c_{i}\},\phi ,v)$ is obtained from the effective action in the
symmetric phase $\Gamma _{SYM}(\{c_{i}\},u,\phi )$ as 
\begin{equation}
\Gamma _{BS}(\{c_{i}\},\phi ,v)=\left[ \Gamma _{SYM}(\{c_{i}\},u,\phi
+v)\right] _{u=v}.
\end{equation}
By applying $D$ to the minimization condition for the VEV given in Eq. (8) and
by using Eq. (14), we can obtain the RG behavior of VEV as \cite{kim2} 
\begin{equation}
Dv=\frac{(\gamma _{\phi }+\widehat{\gamma })v-C_{u}(v)\gamma _{u}}{%
1-C_{u}(v)/v},
\end{equation}
and by applying $D$ to the effective action in the broken symmetry phase $%
\Gamma _{BS}(\{c_{i}\},\phi ,v)$ and by using Eqs. (14),(15) and (18), we
obtain 
\begin{eqnarray}
D\Gamma _{BS}(\{c_{i}\},\phi ,v) &=&\left[ D\Gamma _{SYM}(\{c_{i}\},u,\phi
+v)\right] _{u=v}+(Dv)\left[ \frac{\partial V_{SYM}(\{c_{i}\},u,v)}{\partial
v}+\frac{\partial V_{SYM}(\{c_{i}\},u,v)}{\partial u}\right] _{u=v} 
\nonumber \\
&=&-\gamma _{\phi }(\phi +v)\left[ \frac{\partial \Gamma
_{SYM}(\{c_{i}\},v,\phi +v)}{\partial \phi }\right] _{u=v}-(\widehat{\gamma }%
\text{ }v-C_{u}(v)\gamma _{u})\left[ \Gamma _{SYM}(\{c_{i}\},v,\phi
+v)\right] _{u=v}  \nonumber \\
&+&(Dv)(1-\frac{C_{u}(v)}{v})\left[ \frac{\partial \Gamma
_{SYM}(\{c_{i}\},u,v)}{\partial v}\right] _{u=v}=-\gamma _{\phi }\phi \frac{%
\partial \Gamma _{BS}(\{c_{i}\},\phi ,v)}{\partial \phi }.
\end{eqnarray}
By comparing this equation with that in case of the $R_{\xi }$ gauge given
in Eq. (9), we can see that the RG equation for the effective action in the
broken symmetry phase $\Gamma _{BS}(\{c_{i}\},\phi ,v)$ is the same in both the $%
R_{\xi }$ and the $\overline{R_{\xi }}$ gauges if we substitute $v(\{c_{i}\})$
determined from the minimization condition given in Eq. (8) for the VEV in $%
\Gamma _{BS}(\{c_{i}\},\phi ,v)$ as in the case of the $R_{\xi }$ gauge.

Now, let us consider the running of the pole mass in the broken symmetry
phase in the $R_{\xi }$ and the $\overline{R_{\xi }}$ gauges.\ The pole mass $%
M^{2}$ in the broken symmetry phase is defined as a pole of the two-point
Green's function as 
\begin{equation}
\left[ \frac{\delta ^{2}\Gamma _{BS}(\{c_{i}\},\phi ,v)}{\delta \phi ^{2}}%
\right] _{\phi =0,p^{2}=-M^{2}}=0.
\end{equation}
By taking the derivative $\frac{\delta }{\delta \phi }$ of the RG
equation for the effective action in the broken symmetry phase (Eqs. (9) and
(19)) twice, we obtain the RG equation for $\frac{\delta ^{2}\Gamma _{BS}}{\delta
\phi ^{2}}$ as 
\begin{equation}
D\frac{\delta ^{2}\Gamma _{BS}(\{c_{i}\},\phi ,v)}{\delta \phi ^{2}}+2\gamma
_{\phi }\frac{\delta ^{2}\Gamma _{BS}(\{c_{i}\},\phi ,v)}{\delta \phi ^{2}}%
+\gamma _{\phi }\phi \frac{\delta ^{3}\Gamma _{BS}(\{c_{i}\},\phi ,v)}{%
\delta \phi ^{3}}=0
\end{equation}
in both the $R_{\xi }$ and the $\overline{R_{\xi }}$ gauges. Then by applying $%
D $ to Eq. (20) and by using Eq. (21), we obtain 
\begin{equation}
0=D\left[ \frac{\delta ^{2}\Gamma _{BS}(\{c_{i}\},\phi ,v)}{\delta \phi ^{2}}%
\right] _{\phi =0,\text{ }p^{2}=-M^{2}}=(DM^{2})\left[ \frac{\delta
^{3}\Gamma _{BS}(\{c_{i}\},\phi ,v)}{\delta \phi ^{2}\delta (p^{2})}\right]
_{\phi =0,\text{ }p^{2}=-M^{2}}-\left[ 2\gamma _{\phi }\frac{\delta
^{2}\Gamma _{BS}(\{c_{i}\},\phi ,v)}{\delta \phi ^{2}}\right] _{\phi =0,%
\text{ }p^{2}=-M^{2}}.
\end{equation}
Because the second term of above equation vanishes due to Eq. (20), we
conclude that 
\begin{equation}
DM^{2}=0
\end{equation}
and hence the pole mass is RG invariant in both cases. Finally, we will see
that the pole mass of the Higgs field for both the $R_{\xi }$ and the $%
\overline{R_{\xi }}$ gauges are exactly the same up to one-loop. Because the one-loop
pole mass for the $\overline{R_{\xi }}$ gauge given in Eq. (41) of
Ref. (8) is RG invariant, this shows that the one-loop pole mass for $%
R_{\xi }$ is also RG invariant. In order to see this, let us note that
up to one-loop, the pole mass is given by
\begin{equation}
M^{2}=-2m^{2}+[\Pi ^{(1)}(p^{2})]_{p^{2}=-2m^{2}},
\end{equation}
where the contributions to the one-loop $\Phi _{1}$ self energy $\Pi
^{(1)}(p^{2})$ comes from three types of the diagrams as 
\begin{equation}
\Pi _{XY}=\begin{picture}(50,20) \put(10,5){\line(1,0){8}}
\put(26,5){\circle{16}} \put(34,5){\line(1,0){8}} \put(23,15) {X}
\put(23,-11) {Y} \end{picture},\Pi _{X}=\begin{picture}(50,20)
\put(10,-3){\line(1,0){20}} \put(20,5){\circle{16}} \put(17,15) {X}
\end{picture} ,T_{X}=\begin{picture}(50,20) \put(10,-3){\line(1,0){20}}
\put(20,-3){\line(0,1){5}} \put(20,10){\circle{16}} \put(17,20) {X}
\end{picture}.
\end{equation}
Then, the difference $\Delta M^{2}$ between the one-loop pole masses in the $%
R_{\xi }$ and the $\overline{R_{\xi }}$ gauge is determined by the difference of
the one-loop self energy $\Delta \Pi (p^{2})$ between that of the $R_{\xi }$
gauge ($\Pi ^{R_{\xi }}(p^{2})$) and the $\overline{R_{\xi }}$ gauge ($\Pi ^{%
\overline{R_{\xi }}}(p^{2})$) as
\begin{equation}
\Delta M^{2}=[\Pi ^{R_{\xi }}(p^{2})-\Pi ^{\overline{R_{\xi }}%
}(p^{2})]_{p^{2}=-2m^{2}}\equiv [\Delta \Pi (p^{2})]_{p^{2}=-2m^{2}}.
\end{equation}
By using the given Feynman rules of the $R_{\xi }$ and the $\overline{R_{\xi }}$
gauges given in Ref. 7, we obtain the non-zero contributions to $\Delta \Pi
\equiv \Pi _{R_{\xi }}-\Pi _{\overline{R_{\xi }}}$ as 
\begin{eqnarray}
\lbrack \Delta \Pi _{A\Phi _{2}}(p^{2})]_{p^{2}=-2m^{2}} &=&(\frac{2}{3}\xi
\lambda -\xi ^{2}g^{2})m_{A}^{2}[B(p^{2},\xi m_{A}^{2},\xi
m_{A}^{2})]_{p^{2}=-2m^{2}}+\xi g^{2}A(\xi m_{A}^{2}),  \nonumber \\
\lbrack \Delta \Pi _{\Phi _{2}\Phi _{2}}(p^{2})]_{p^{2}=-2m^{2}} &=&(-\frac{%
2}{3}\xi \lambda -2\xi ^{2}g^{2})m_{A}^{2}[B(p^{2},\xi m_{A}^{2},\xi
m_{A}^{2})]_{p^{2}=-2m^{2}},  \nonumber \\
\lbrack \Delta \Pi _{\overline{c}c}(p^{2})]_{p^{2}=-2m^{2}} &=&3\xi
^{2}g^{2}m_{A}^{2}[B(p^{2},\xi m_{A}^{2},\xi
m_{A}^{2})]_{p^{2}=-2m^{2}},  \nonumber \\
\lbrack \Delta \Pi _{\Phi _{2}}]_{p^{2}=-2m^{2}} &=&\xi g^{2}A(\xi
m_{A}^{2}),  \nonumber \\
\lbrack \Delta \Pi _{c}]_{p^{2}=-2m^{2}} &=&-2\xi g^{2}A(\xi m_{A}^{2})\text{
},  \nonumber \\
\lbrack \Delta T_{\Phi _{2}}]_{p^{2}=-2m^{2}} &=&-\xi m_{A}^{2}A(\xi
m_{A}^{2})\text{ },  \nonumber \\
\lbrack \Delta T_{c}]_{p^{2}=-2m^{2}} &=&\xi m_{A}^{2}A(\xi m_{A}^{2})\text{ 
},
\end{eqnarray}
where $A(m^{2})$ and $B(p^{2},m_{1}^{2},m_{2}^{2})$ are the one-loop
functions introduced by Passarino and Veltman as \cite{Passarino} 
\begin{eqnarray}
A(m^{2}) &=&\int \frac{d^{D}q}{(2\pi )^{D}}\frac{1}{q^{2}+m^{2}},  \nonumber
\\
B(p^{2},m_{1}^{2},m_{2}^{2}) &=&\int \frac{d^{D}q}{(2\pi )^{D}}\frac{1}{%
(q^{2}+m_{1}^{2})((p+q)^{2}+m_{2}^{2})},
\end{eqnarray}
and the mass of the gauge boson $m_{A}^{2}$ is given by $g^{2}v^{2}.\ $In
order to obtain a consistent one-loop result, we have used the tree-level
value for the VEV as $v^{2}=-\frac{6m^{2}}{\lambda }.$ By summing all the terms
given in Eq. (27), we obtain $[\Delta \Pi (p^{2})]_{p^{2}=-2m^{2}}=0$
Hence, $\Delta M^{2}=0\ $ (Eq. (26)) so that the pole masses in the $R_{\xi }$
and the $\overline{R_{\xi }}$ gauge are the same up to one-loop.

In conclusion, we have shown that although the RG running of the effective
potential in the symmetric phase is quite different in the $R_{\xi }$ and the $%
\overline{R_{\xi }}$ gauges and gives a different result for the running of
the VEV, if we substitute $v(\{c_{i}\})$ determined from the minimization
condition given in Eq. (8) for the VEV in $\Gamma _{BS}(\{c_{i}\},\phi ,v)$,
then the effective potential in the broken symmetry phase satisfies the same
RG equation as that of the RG function obtained in the symmetric phase. Also, the
pole mass is RG invariant in the broken symmetry phase of both gauges and
is exactly the same in both gauges at one-loop order.

\end{document}